\documentclass{appolb}
\usepackage{epsfig}

\begin{document}
\title{Restoration of singlet axial symmetry at finite temperature%
\thanks{Presented at XXVIII Max Born Symposium, Wroclaw, 19-21 May, 2011}%
}
\author{S. Beni\' c
\address{Physics Department, Faculty of Science,
University of Zagreb,
Bijeni\v{c}ka c. 32, Zagreb 10000, Croatia}
\and
D. Horvati\'c$^a$, D. Kekez$^b$, D. Klabu\v car$^a$
\address{$^a$Physics Department, Faculty of Science,
University of Zagreb,
Bijeni\v{c}ka c. 32, Zagreb 10000, Croatia
\\
$^b$Rugjer Bo\v skovi\'c Institute, 
Bijeni\v{c}ka c. 54, Zagreb 10000, Croatia}
}
\maketitle
\begin{abstract}
To accomodate recent RHIC data on $\eta'$ multiplicity, we propose a minimal modification of the Witten-Veneziano relation at high temperature. This renders a significant drop of $\eta'$ mass at high temperature signaling a restoration of the $U(1)_{A}$, and the Goldstone character of $\eta'$. 
\end{abstract}
\PACS{11.10.St, 11.10.Wx, 12.38.-t, 12.38.Aw, 24.85.+p}
  
\section{Introduction}

\begin{figure}
\begin{center}
\includegraphics[scale=0.8]{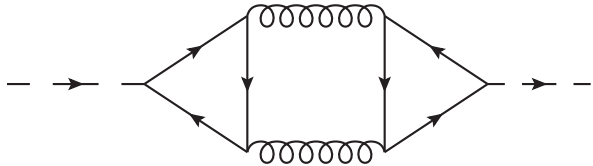}
\end{center}
\caption{The lowest order contribution to the excess of the $\eta'$ mass.}
\label{diam}
\end{figure}

On the classical level, QCD with $N_f$ massless flavors enjoys a global chiral $U(N_f)_L\otimes U(N_f)_R$ symmetry. The subgroup with unit determinant, namely the $SU(N_f)_L\otimes SU(N_f)_R$, is spontaneously broken down to its diagonal part $SU(N_f)_{L+R}=SU(N_f)_V$. This gives rise to $8$ pseudoscalar Goldstone bosons: $\pi$'s, $K$'s and the $\eta$.\\
On the other hand, the non-abelian anomaly in the $U(1)_A$ sector,
\begin{equation}
\partial_{\mu}J_{5\mu}^{0}(x)=\frac{g_{s}^2}{32\pi^{2}}\epsilon^{\mu\nu\rho\sigma}F^{\mu\nu}_{a}F^{\rho\sigma}_{a}
\end{equation}
explicitely breaks $U(1)_A$ symmetry. Coupled with non-trivial topology in the field space of QCD, it prevents the spontaneous breaking of $U(1)_{A}$, thus generating the extra mass for $\eta'$. The simplest example of such contribution to $m_{\eta'}$ is the diamond diagram in Fig. \ref{diam}.\\
Nevertheless, the fate of $U(1)_{A}$ at $T>0$ could be drastically changed. New RHIC data from central Au+Au collisons \cite{Csorgo:2009pa} on $\eta'$ multiplicity shows a drop in it's mass by at least $200$ MeV inside the fireball
\begin{equation}
\label{exp}
m_{\eta'}(\mathrm{vacuum})=958\mbox{ MeV}\rightarrow m_{\eta'}(\mathrm{fireball})=340^{+50+280}_{-60-140}\pm 45\mbox{ MeV},
\end{equation}
thus signaling restoration of the Goldstone character of $\eta'$.

\section{Witten-Veneziano relation at finite $T$}

On the theoretical side, excess in the $\eta'$ mass is inferred from the Witten-Veneziano relation (WVR) \cite{Witten:1979vv,Veneziano:1979ec}
\begin{equation}
\label{wvr}
M_{\eta}^{2}+M_{\eta'}^{2}-2M_{K}^{2}=\frac{2N_{f}}{f_{\pi}^{2}}\chi_{\mathrm{YM}}.
\end{equation}
Due to the flavor content of the pseudoscalars, WVR just says that the excess in the singlet $\eta¢^0$ mass is coming from the glue sector, i.e. $\chi_{\mathrm{YM}}$ is the Yang-Mills topological susceptibility.\\ 
Previous studies \cite{Horvatic:2007qs} of WVR indicated, for various lattice forms of $\chi_{\mathrm{YM}}$, that $\eta'$ mass increases as $\chi_{\mathrm{YM}}$ melts, \textit{in contrast} to the result (\ref{exp}). This is happening as $T$ approaches $T_\mathrm{Ch}$, due to the fact that there $f_\pi(T)$ starts decreasing significantly.\\
\begin{figure}
\begin{center}
\includegraphics[scale=0.4]{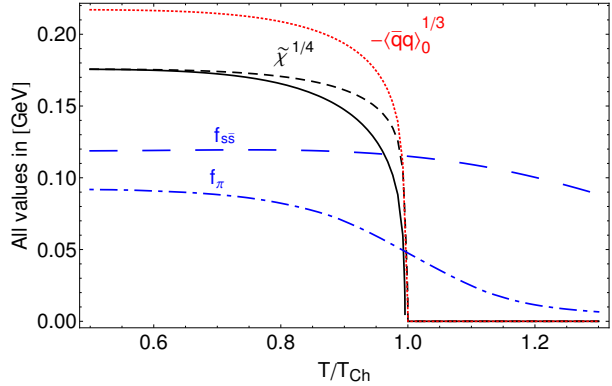}
\end{center}
\caption{The relative-temperature dependences, on $T/T_{\rm Ch}$, of
${\widetilde \chi}^{1/4}$, $\langle{\bar q}q\rangle_0^{1/3}$,
$f_\pi$ and $f_{s\bar s}$. The solid curve depicts ${\widetilde \chi}^{1/4}$ for $\delta=0$, and the short-dashed curve is
${\widetilde \chi}^{1/4}$ for $\delta=1$.  At $T=0$, both
${\widetilde \chi}\,$'s are equal to $\chi_{\rm Y\!M} = (0.1757 
\, \rm GeV)^4$, the weighted average \cite{Horvatic:2007qs}
of various lattice results for $\chi_{\rm Y\!M}$.}
\label{temp1}
\end{figure}

\subsection{Connection between QCD and YM topological susceptibility}

In that light, we propose \cite{Benic:2011fv} a minimal modification of WVR first by using Leutwyler-Smilga result \cite{Leutwyler:1992yt} in the vacuum
\begin{equation}
\label{ls}
\chi_{\mathrm{YM}}=\frac{\chi}{1+\chi\frac{N_{f}}{\Sigma m}}.
\end{equation}
Just like the WVR, Leutwyler-Smilga relation connects the quantities from two different theories: $\chi$ is the QCD topological susceptibility, and Eq. (\ref{ls}) shows that it approaches $\chi_{\mathrm{YM}}$ only for large quark masses whereas $\chi_{\mathrm{YM}}$ and $\chi$ are very different for light quarks. $\Sigma=\langle\bar{q}q\rangle_{0}$ is the condensate in the chiral limit.\\
We denote by $\tilde{\chi}$ the whole right hand side of Eq. (\ref{ls}), and use it at finite temperature.\\
Importance of this relation comes from the fact that $\chi$ is driven by the chiral quark condensate in the leading order of expansion in small quark masses
\begin{equation}
\label{dvv}
\chi=-\frac{\Sigma m}{N_{f}}+C_{m},\quad \frac{N_{f}}{m}=\sum_{f}\frac{1}{m_{f}}.
\end{equation}
This is the di Vecchia-Veneziano result \cite{Leutwyler:1992yt, Di Vecchia:1980ve}. The next term in this expansion, $C_{m}$, is essenital as it keeps $\chi_{\mathrm{YM}}$ from blowing up. We fix its value at $T=0$ by demanding $\tilde{\chi}(0)=\chi_{\mathrm{YM}}$.

\subsection{Exploring the thermal dependence}

For the thermal dependence of $\chi$ we use an ansatz
$$\chi(T)=-\frac{\Sigma(T)m}{N_{f}}+C_{m}(0)\Bigl[\frac{\Sigma(T)}{\Sigma(0)}\Bigr]^{\delta}.$$
This gives
\begin{equation}
\label{tildechi}
\tilde{\chi}(T)=\frac{\Sigma(T)m}{N_{f}}\Bigl\{1-\frac{1}{C_{m}(0)}\frac{\Sigma(T)m}{N_{f}}\Bigl[\frac{\Sigma(0)}{\Sigma(T)}\Bigr]^{\delta}\Bigr\}
\end{equation}
The interesting window for $\delta$ is then $0<\delta<1$ since the lower limit gives no thermal dependence for the correction term, and a quadratic one in the last equation for $\tilde{\chi}$. With $\delta=1$ there is an enhancement of the $\eta'$ mass \cite{Benic:2011fv}, and therefore this is the upper limit of interest.\\

\begin{figure}
\begin{center}
\includegraphics[scale=0.4]{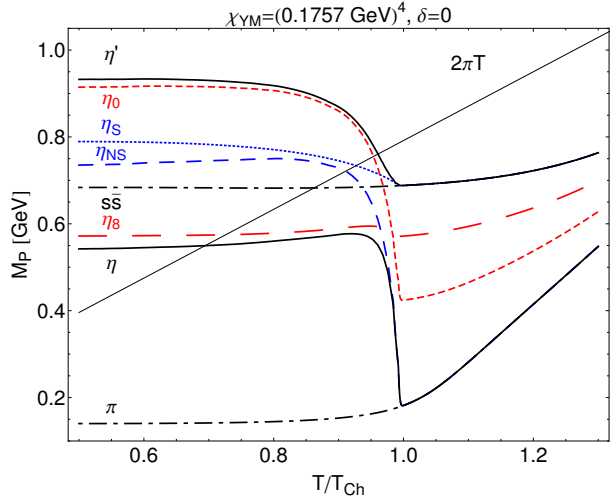}
\end{center}
\caption{The relative-temperature dependences, on $T/T_{\rm Ch}$, of the pseudoscalar masses for $\delta=0$.}
\label{massplot}
\end{figure}

The interesting window for $\delta$ is then $0 \leq \delta \leq 1$, since $\delta = 0$ gives the $T$-independent correction term, while $\delta = 1$ already leads to precursors of the unwanted mass enhancement in the $\eta'-\eta$ complex. Therefore, $\delta \sim 1$ is the upper limit of interest, although the results for the $T$-dependence of the meson masses is quantitatively not much different for $\delta = 1$ than from the case $\delta = 0$, as it is depicted in Figs. \ref{massplot} and \ref{massplot2} respectively.

\section{Results and discussion}

\begin{figure}
\begin{center}
\includegraphics[scale=0.4]{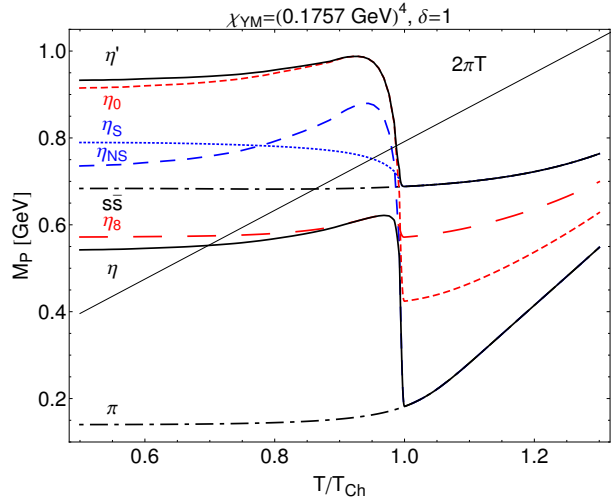}
\end{center}
\caption{The relative-temperature dependences, on $T/T_{\rm Ch}$, of the pseudoscalar masses for $\delta=1$.}
\label{massplot2}
\end{figure}

Mesons are constructed as $q\bar{q}$ bound states via the Bethe-Salpeter equation in the ladder approximation. Dynamical quarks are build up from the Dyson-Schwinger equation in rainbow approximation. We use the succesfull rank-2 separable model \cite{Blaschke:2000gd} for the gluon propagator, which was also used in Ref. \cite{Horvatic:2007qs}. Rainbow-ladder approximation is the simplest symmetry preserving truncation, necessary for the correct chiral behavior of the theory.\\
The bound state approach enables access only to the non-anomalous part of the meson masses, since the ladder Bethe-Salpeter kernel does not include diagrams like in Fig. \ref{diam}. Therefore, the anomalous part is inferred from Eq. (\ref{wvr}).
The strategy is to use flavor mass matrices to extract $\eta$ and $\eta'$ masses from the calculated non-anomalous sector. This is presented in detail in Ref. \cite{Horvatic:2007qs}.\\
Our main result is presented in Figs. \ref{massplot} and \ref{massplot2}. The reduction in the $\eta'$ mass is around $200$ MeV, which is in quantitative agreement with RHIC data. This is possible only due to the proposed modification of the WVR relation at finite $T$. Topological susceptibility of pure glue, $\chi_{\mathrm{YM}}$, is just too resistant, with the characteristic melting temperature being $T_{\mathrm{YM}}=260$ MeV (see e. g. \cite{Alles:1996nm,Boyd:1996bx}).\\
In contrast, the pseudocritical temperatures for the chiral and deconfinement transitions in the full QCD are lower than $T_{\rm YM}$ by some 100 MeV or more (e.g., see Ref. \cite{Fodor:2009ax}) 
due to the presence of the quark degrees of freedom. In that regard, Eq. (\ref{dvv}) is essential as it couples chiral restoration to $U(1)_{A}$ restoration, allowing for $\tilde{\chi}(T)$ to melt away even sooner than $f_{\pi}(T)$. In the separable model used here we have $T_{\mathrm{Ch}}=128$ MeV which is admittedly lower than the so far accepted value around $160$-$170$ MeV. It has been shown \cite{Horvatic:2010md} that the same model coupled with the gluon degrees of freedom in the form of the Polyakov loop cures this discrepancy.\\

\end{document}